\documentclass[usenatbib]{mn2e}
\usepackage{graphicx}
\usepackage{epstopdf}
\usepackage{graphics}
\usepackage{rotating}
\usepackage{color}
\usepackage{amsmath, esint}
\usepackage{bm}
\usepackage{threeparttable}
\usepackage{url}

\bibpunct{(}{)}{;}{a}{}{,}

\begin{document}

\def\HI{H{\sc i} }
\def\HII{H{\sc ii} }
\def\Ha{{\rm H}\alpha }
\def\Msun{{\mathrm M}_{\odot} }
\def\kms{\,km\,s^{-1}}
\def\Msunrm{{\mathrm M}_{\odot} }
\def\Lsunrm{{\mathrm L}_{\odot} }

\title{Tides or dark matter sub-halos: Which ones are more attractive?}

\author[Sylvia Ploeckinger]
{Sylvia~Ploeckinger$^1$\thanks{email: ploeckinger@strw.leidenuniv.nl}\\
	$^1$Leiden Observatory, Leiden University, PO Box 9513, 2300 RA Leiden, The Netherlands }

\maketitle

\begin{abstract}
	
Young tidal dwarf galaxies (TDGs) are observed in the tidal debris of gas-rich interacting galaxies. 
In contrast to what is generally assumed to be the case for isolated dwarf galaxies, TDGs are not embedded in their own dark matter (DM) sub-halo. Hence, they are more sensitive to stellar feedback and could be disrupted on a short time-scale. Detailed numerical and observational studies demonstrate that isolated DM-dominated dwarf galaxies can have lifetimes of more than 10 Gyr. For TDGs that evolve in a tidal field with compressing accelerations equal to the gravitational acceleration within a DM sub-halo typical of an isolated dwarf galaxy, a similar survival time is expected. The tidal acceleration profile depends on the virial mass of the host galaxy and the distance between the TDG and its host. We analytically compare the tidal compression to the gravitational acceleration due to either cuspy or cored DM sub-halos of various virial masses. For example, the tidal field at a distance of 100 kpc to a host halo of $10^{13}\,\Msun$ can be as stabilizing as a $10^9\,\Msun$ DM sub-halo. By linking the tidal field to the equivalent gravitational field of a DM sub-halo, we can use existing models of isolated dwarfs to estimate the survivability of TDGs. We show that part of the unexpectedly high dynamical masses inferred from observations of some TDGs can be explained by tidal compression and hence TDGs require to contain less unobservable matter to understand their rotation curves.

\end{abstract}


\section{Introduction}

In the current standard cosmological theory, $\Lambda$-CDM, where the Universe is dominated by dark energy ($\Lambda$) and galaxies by their cold (C) dark matter (DM) content, dwarf galaxies (DGs) are the remainders of the hierarchical structure formation process. They form the low-mass end of DM-dominated objects, as they have not yet merged into more massive systems. State-of-the-art cosmological simulations, such as Illustris \citep{Vogelsberger2014}, EAGLE \citep{Schaye2014c} and zooms of Local Group analogues with the EAGLE code \citep{Sawala2015} can now follow the evolution of hundreds or thousands of DGs in one simulation box with increasingly high resolution. The stellar feedback processes that are especially important to regulate star formation (SF) in low-mass galaxies, can be calibrated to observed relations, such as the galaxy stellar mass function. However not all discrepancies between simulations and observations on the scale of DGs are solved \citep{Kroupa2015}. 

The simulations complement deep surveys of dwarf galaxies in the Local Group \citep{Holtzman2006,Orban2008}, the local volume  \citep{Hunter2012, Lee2009} or in the Virgo \citep{Lisker2007,DeLooze2013,Toloba2015} and Fornax clusters \citep{Mieske2007,Chilingarian2011}. Both observations and simulations indicate that very extended ($>10\,\mathrm{Gyr}$) lifetimes are possible for isolated DGs \citep[e.g.][]{Sawala2015, Karachentsev2014, Karachentsev2015}. The explanation for their long-time survival is usually found in the DM sub-halo that engulfs the DG, increases the binding energy and stabilizing it against disrupting processes such as stellar feedback. 

Tidal dwarf galaxies (TDGs) form in the tidal debris of interacting gas-rich galaxies.
In $\Lambda$-CDM, the merging host galaxies are surrounded by a DM halo, but the velocity dispersion of these DM particles is too high to be captured by newly formed, low-mass TDGs which are therefore not expected to be hosted by DM sub-halos \citep{Barnes1992, Bournaud2006, Wetzstein2007}. Observations show that TDGs have active SF \citep{Braine2000, Braine2001} and hence stellar feedback could disrupt the TDG on a short time-scale without a DM sub-halo to keep it stable. Old TDGs are difficult to identify, as their prominent birth-place, the tidal arm, typically dissolves on a time-scale of several hundred Myr \citep[e.g.][]{Hibbard1995,Bournaud2006}. Without detailed kinematic or spectroscopic studies, old TDGs might be hard to distinguish from other stellar objects in the same mass range. Only a few simulations of the evolution of individual TDGs  \citep[e.g.][]{Recchi2007, Smith2013, Yang2014, Ploeckinger2014, Ploeckinger2015} or their formation during galaxy interactions \citep[e.g.][]{Wetzstein2007,Bournaud2003,Bournaud2006,Hammer2010,Fouquet2012} are available in the literature.

While cosmological simulations are beginning to resolve the mass scales relevant for TDG formation, the feedback efficiencies are sometimes taken to be functions of the properties of the DM \citep[e.g.][]{Vogelsberger2014}, which is not a good strategy for the stellar feedback inside TDGs. If many TDGs survive for several Gyr, then the number of old TDGs hiding in the DG and super-star cluster populations could be large. The first observational evidence for their long-term survival was found in a survey around early-type galaxies, including the detection of the oldest TDG identified so far, with an age of around 4 Gyr \citep{Duc2014}.  

In \citet{Ploeckinger2015} the evolution of individual TDGs is simulated for 3 Gyr. During their evolution, the TDGs pass both the apo- and the peri-centres of their eccentric orbits around their host galaxy. While they are close to the apo-centre, the star formation self-regulates to a moderate level, but the TDGs become  significantly tidally compressed as they get near the host galaxy \footnote{For a movie of the tidal compression of model \mbox{TDG-p} from \citet{Ploeckinger2015}, see \url{https://sites.google.com/site/sylviaploeckinger/tidal-dwarf-galaxies/analytical}}. Unlike star clusters, where tidal compression leads to dynamical heating of the stellar system, TDGs are gas-rich and dissipative processes are important for their formation \citep{Wetzstein2007}. The tidal field compresses the gas phase, leading to efficient cooling and a high star formation rate. In addition, this compression helps to stabilize the TDG against the subsequent stellar feedback \citep{Ploeckinger2015}. \citet{Kroupa1997} and \citet{Klessen1998} have demonstrated that gas-free TDGs can survive for 10 Gyr despite repeated tidal harassing.

Tidal forces are often used to study the disruption of objects (such as stars close to a black hole), but a tidal field can also trigger the formation of new systems or increase their stability in its compressive mode \citep{Renaud2009,Renaud2014}. In regimes where the compression of the tidal field dominates, an object experiences an extra acceleration towards its centre, reminiscent of an additional gravitational acceleration as e.g. caused by a DM sub-halo. By comparing the acceleration due to the tidal field to that caused by a DM sub-halo of a given mass, we can relate the evolution of TDGs to that of their much better studied isolated DG siblings: If an isolated DM-dominated DG of a given mass can survive its stellar feedback for almost a Hubble time, then a similar TDG at a position within a tidal field with comparable accelerations should survive as well. 

We present a fully analytic method to assign an equivalent DM sub-halo mass to a TDG depending on the acceleration due to the tidal field and therefore on the mass and profile of their host halo and the distance between the TDG and the centre of the host galaxy. The method provides a simple tool to predict or explain the survival of TDGs. \citet{Bournaud2007} measured the rotation curves of three TDGs in NGC 5291 and found the ratio between the dynamical and the visible mass to have a value of 2 to 3. As TDGs should be free of DM, these ratios are unexpectedly high. We show that the compression of the tidal field can lead to a comparable increase of the dynamical mass as a DM-halo with a mass equal to the ``missing mass" in these TDGs could. Therefore this method can help to explain an apparent DM content in TDGs. 

In Sec.~\ref{Sec:tidalfields} the general properties of tidal fields are summarized and applied to find distance limits for each host halo mass, in which the tidal compression along two directions dominates the tidal stretching in the perpendicular direction. In Sec.~\ref{Sec:ayar} the accelerations from the tidal compression are compared to those inside DM sub-halos. In Sec.~\ref{Sec:application} we present two applications of this method: explaining the age of the old TDG in NGC 5557 and investigating the high dynamical masses of the TDGs in the gaseous ring around NGC 5291. We discuss the limitations of the proposed analytic method in Sec.~\ref{Sec:discussion}.

\section{Tidal fields} \label{Sec:tidalfields}

\begin{figure}
	\includegraphics[width = \linewidth]{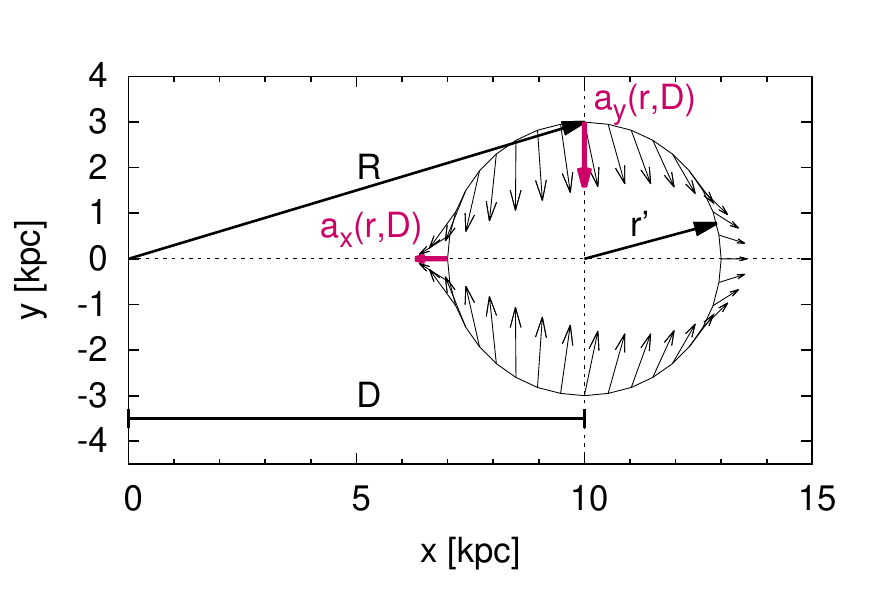}
	\caption{The vectors show the tidal field at a distance of $D = 10 \,\mathrm{kpc}$ to the centre of an NFW potential at $x = y = 0$. Throughout the paper the definitions for $x$, $y$, $D$, $r=|\bm{r'}|$, $\bm{R}$, $a_x$, and $a_y$ are used as indicated here.}
	\label{fig:tides_general}
\end{figure}

For the method presented here, we consider a host galaxy centred at $x=y=z=0$ with a spherically symmetric gravitational potential $\Phi_{\mathrm{host}}$ from a mass distribution $\rho_\mathrm{host}$ and investigate the acceleration field at the surface of a test sphere with radius $r$ at a distance $D$, as shown in Fig.~\ref{fig:tides_general}. 

The acceleration field in the reference frame of $r = 0$ (tidal field), caused by the external potential is described as:

\begin{equation} \label{Eq:tidal}
\frac{\mathrm{d}^2 \bm{r}'}{\mathrm{d}t^2} = \nabla \Phi_{\mathrm{host}}(\bm{0}) - \nabla \Phi_{\mathrm{host}} (\bm{r}')
\end{equation}

\noindent
where $\bm{r}'$ is the position vector relative to the centre of the test sphere. The gravitational potential of the host galaxy is described by an NFW profile \citep{Navarro1996} of

\begin{equation}
\rho_{\mathrm{NFW}}(R) = \frac{\rho_{0,h}}{\frac{R}{r_{0,h}} \left ( 1+\frac{R}{r_{0,h}}\right )^2}	
\end{equation}

\noindent
with the resulting potential

\begin{equation}\label{Eq:NFW1}
	 \Phi_{\mathrm{NFW}} (R) = - \frac{G \rho_{0,h} r_{0,h}^3}{R} \cdot \log(1+R/r_{0,h})
\end{equation}

\noindent
where $\rho_{0,h}$ is the characteristic density of the host halo and $r_{0,h}$ its scale radius. The tidal field at $z=0$ for a test sphere of $r$ = 3 kpc at a distance of $D$ = 10 kpc to the centre of an NFW halo with $M_{\mathrm{vir}} = 10^{12}\,\Msun$ is indicated as vectors in Fig.~\ref{fig:tides_general} for illustration. In this case, the test sphere would be stretched in $x$-direction and compressed along the $y$- and $z$ axis. In a potential with a central cusp, such as the NFW profile, the tidal field is never compressive on all axes \citep[see e.g.][]{Renaud2009}. Nevertheless, it is possible to look for regimes, where the compression dominates the expansion, which is the aim of this section. Note that during a galaxy interaction, the gravitational potential will not be relaxed. In principle, the method can be used for arbitrary gravitational potentials if the analytical are replaced by the numerical derivations. For simplicity, we show the idealised case of an NFW potential, but see the discussion on this in Sec.~\ref{Sec:discussion}.

\subsection{Gauss's law of gravity}

The net flow  over the whole surface of the test sphere, shown in Fig.~\ref{fig:tides_general} is directly given by the integral form of Gauss's law for gravity:

\begin{equation} \label{Eq:gauss}
	\oint_{\partial V} \bm{g} \cdot \mathrm{d}\bm{A} = - 4 \pi G M_{\mathrm{enc}} \quad .
\end{equation}

\noindent
It states, that the integral of all accelerations $\bm{g}$ in a gravitational potential over any static surface of the volume $\partial V$ is proportional to its enclosed mass $M_{\mathrm{enc}}$. For the above example this means, that if the massless test sphere is outside of the density distribution $\rho_\mathrm{host}$ and therefore $M_{\mathrm{enc}} = 0$, the net effect over the surface elements $\bm{A}$ vanishes. The tidal stretching in some directions is therefore exactly balanced by the tidal compression in other directions. If the satellite object is inside the mass distribution, then $M_{\mathrm{enc}} > 0$ and subsequently the net effect on a static surface is always compressing ($\oint_{\partial V} \bm{g} \cdot \mathrm{d}\bm{A} < 0$).

Therefore, whenever a TDG is inside a DM host halo, there is a net compression caused by the tidal field. In a related work, \citet{Bournaud2003} showed that massive TDGs at the tips of tidal arms only form in their simulations if the DM halo of the host is not truncated. Outside the truncation radius, the tidal field doesn't lead to the necessary compression for the formation of TDGs. 

A first condition for the formation and survival of TDGs is therefore to reside within the host galaxy's DM halo which translates to $D<r_{\mathrm{vir}}$, where $r_{\mathrm{vir}}$ is the virial radius. In addition, the net compression over a surface is proportional to the enclosed mass within it. A decreasing distance to the host galaxy leads to an increasing enclosed DM mass and therefore to a higher net compression.

\subsection{Tidal tensor}

\begin{figure}
	\includegraphics[width = \linewidth]{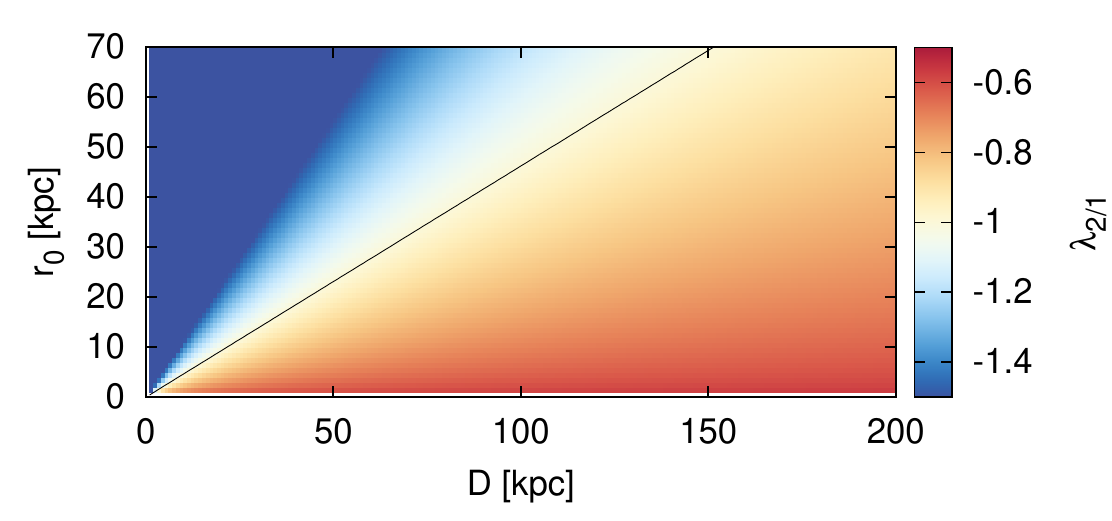}
	\caption{The colour scale represents the ratio between the eigenvalues of the tidal tensor $\lambda_{2/1}$ at distances D (x-axis) to the centre of NFW potentials with scale radius $r_0$ (y-axis). The solid black line shows where $\lambda_{2/1} = -1$ (see Eq.~\ref{Eq:lambda}). The blueish region above the solid line corresponds to a contraction dominated regime $(\lambda_{2/1} < -1)$, while in the reddish region below the black line, the expansion dominates $(\lambda_{2/1} > -1)$. For more massive halos and therefore NFW potentials with larger scale radii $r_0$, the region in which the compression dominates the expansion extends out to larger distances. For more information see text.}
	\label{fig:lambda21}
\end{figure}

\begin{figure*}
	\begin{minipage}{0.6\linewidth}
		\includegraphics[width = \linewidth]{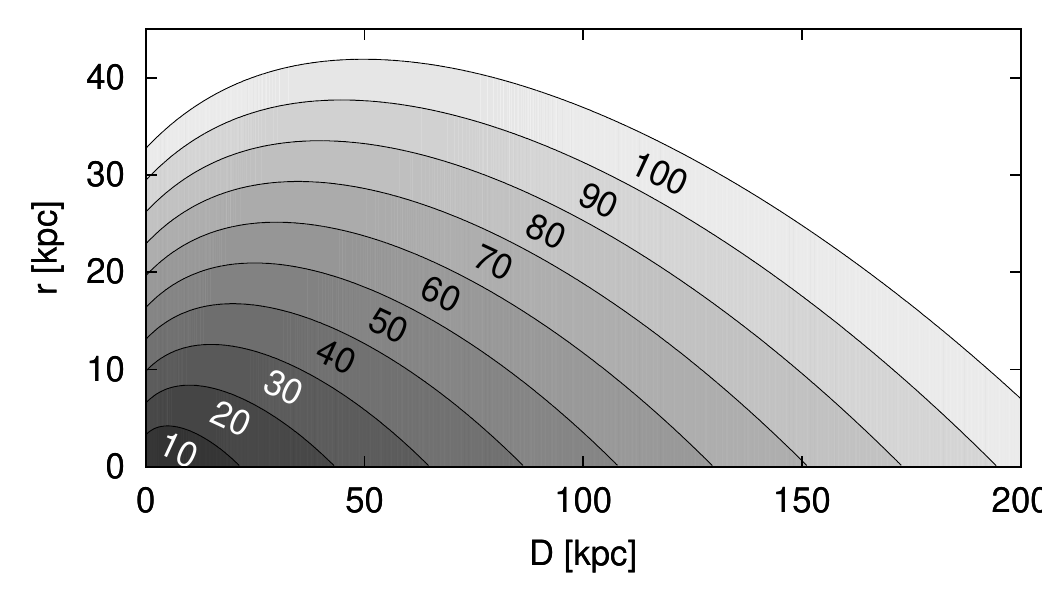}
	\end{minipage}
	\begin{minipage}{0.3\linewidth}
		\begin{center}
			\begin{tabular}{cc}
				\hline
				\hline
				$r_0$			&	 $M_{\mathrm{vir}}$	\\
				$[$kpc$]$ 		&	log $[M / \Msun]$   \\		
				\hline
				10  &   11.11  \\
				20  &   11.83  \\
				30  &   12.25  \\
				40  &   12.55  \\
				50  &   12.78  \\
				60  &   12.96  \\
				70  &   13.12  \\
				80  &   13.25  \\
				90  &   13.37  \\
				100  &   13.47  \\
				\hline
			\end{tabular}
		\end{center}
	\end{minipage}
	\caption{The ratio between the tidal acceleration in y- and x-direction ($a_{y/x}$, see also Fig.~\ref{fig:tides_general}) is presented in dependence on the distance to the host galaxy (x-axis),the radial extent of the TDG (y-axis), and the scale radius $r_0$ of the host halo (solid lines). All lines show $a_{y/x} = 1$ for NFW potentials with different scale radii according to Eq.~\ref{Eq:ayax} and are labelled with the value for $r_0$ in kpc. The grey regions indicate where $a_{y/x} > 1$ and therefore the compression dominated region for each scale radius. The corresponding virial mass $M_{\mathrm{vir}}$ for a halo with scale radius $r_0$ is listed in the table on the right (see Sec.~\ref{Sec:ayax} for details). }
	\label{Fig:ayax}
\end{figure*}

One way to describe the tidal field is with the tidal tensor $\bm{T}_t^{ij} \equiv - \partial_i (\partial_j \Phi_{\mathrm{host}})$. $\bm{T}_t^{ij}$ is a symmetric tensor and in its orthogonal form, the eigenvalues $\lambda$ represent the strength and direction of the tidal field along the corresponding eigenvectors $\bm{v}$. If $\lambda_i < 0$, the tidal acceleration leads to a compression in the direction of $\bm{v}_i$ while $\lambda_i > 0$ represents an expansion along $\bm{v}_i$.

In the geometry illustrated in Fig.~\ref{fig:tides_general}, the components of the tidal tensor at a distance $D$ to the host galaxy are defined as:

\begin{eqnarray} \label{Eq:tensor}
\bm{T}^{11}(D) 		&=& - \left (\frac{\partial^2 \Phi_{\mathrm{host}}(D)}{\partial x^2} \right )	\\
\bm{T}^{22}(D) 		&=& - \left (\frac{\partial^2 \Phi_{\mathrm{host}}(D)}{\partial y^2} \right )	\\
\bm{T}^{33}(D) 		&=& - \left (\frac{\partial^2 \Phi_{\mathrm{host}}(D)}{\partial z^2} \right )	\\
\bm{T}^{i \neq j}	&=& 0 
\end{eqnarray}

\noindent
with the eigenvalues $\lambda_1 = \bm{T}^{11}$,  $\lambda_2=\bm{T}^{22}$,  and $\lambda_3=\bm{T}^{33}$) and the eigenvectors $\bm{v}_1$ = (1,0,0), $\bm{v}_2$ = (0,1,0), and $\bm{v}_3$ =(0,0,1). 

The trace of the tidal tensor can then be directly linked to Poisson's law:

\begin{equation} \label{Eq:poisson}
\sum_{i = 1}^{3} \lambda_i = \sum_{i = 1}^{3} - \partial_i (\partial_i \Phi_{\mathrm{host}} ) = - \nabla^2 \Phi_{\mathrm{host}} =  - 4 \pi G  \rho \quad .
\end{equation}

\noindent
As the density $\rho$ is always positive there is no solution where all $\lambda_i$ are positive and where therefore the tidal field would lead to an expansion along all eigenvectors. Depending on $\Phi_{\mathrm{host}}$ and the distance $D$, between one and three eigenvalues can be negative and therefore compressing the test sphere in the corresponding direction. In the example shown in Fig.~\ref{fig:tides_general} the eigenvalue in x-direction is positive while those in y- and z-direction are negative. Fully compressive tidal modes where all $\lambda_i < 0$ appear in central regions of cored potentials or in the overlay of two potential e.g. as in interacting galaxies. For a detailed description of the tidal tensor and a discussion on the different modes of a tidal field, see \citet{Renaud2009}.

Even if the tidal mode is not fully compressive, it's effect can still be more stabilizing than disrupting. 
An indication for that is the ratio between $\lambda_1$ and $\lambda_2$. The fully compressive mode is then described when all $\lambda_i < 0$ while for $\lambda_2 / \lambda_1 < -1$ the contraction in y-direction is stronger than the expansion in x-direction. Note that for the geometry used here, the problem is symmetric around the x-axis and therefore $\lambda_3 = \lambda_2$. 

For an NFW potential (Eq.~\ref{Eq:NFW1}) the ratio $\lambda_{2/1} = \lambda_2 / \lambda_1$ can be calculated analytically\footnote{The Mathematica notebook for the derivation and further use of Eqs. \ref{Eq:lambda},  \ref{Eq:ayax}, and \ref{Eq:ayar} as well as an illustration for Eq.~\ref{Eq:ay} can be downloaded at \url{https://sites.google.com/site/sylviaploeckinger/tidal-dwarf-galaxies/analytical}} and depends only on $D$ and the scale radius $r_0$:

\begin{equation} \label{Eq:lambda}
\begin{split}
&	\lambda_{2/1} (D, r_0) = \\
&	\frac{D^2+D r_0-\left(D+r_0\right)^2 \log \left(\frac{D+r_0}{r_0}\right)}{2 \left(D^2+r_0 (2 D+r_0)\right) \log \left(\frac{D+r_0}{r_0}\right)-3 D^2-2 D r_0}
\end{split}
\end{equation}

\noindent
In Fig.~\ref{fig:lambda21} the relation between $D$, $r_0$ and $\lambda_{2/1}$ is illustrated. The black solid line indicates where the contraction in y- (and z-) direction is equal to the expansion in x-direction. Above the $\lambda_{2/1} = -1$ line, the contraction is larger than the expansion, leading to a first limit on where the survivability of objects could be enhanced rather than decreased.

\subsection{Tidal accelerations} \label{Sec:ayax}

The tidal field at the surface of the test sphere does not only depend on the distance to and the shape of the gravitational potential of the host, but also on its radial extent $r$. As it can be directly seen from Eq.~\ref{Eq:tidal}, the more extended the test sphere is, the larger is the difference between the acceleration at its mass centre and those at its surface. 

In addition to the dependence of the absolute values of the tidal forces on the radius $r$, also the ratio between compression and expansion changes with $r$. In order to include the radial dependence, we calculate the ratio $a_{y/x} = | a_y / a_x |$. We define $a_y$ as the acceleration along the y-axis at the point ($x = D$, $y = r$, $z = 0$; see Fig.~\ref{fig:tides_general}) and use it as a proxy for the maximum compression. Analogously, $a_x$ presents the acceleration in x-direction at the point ($x = D-r$, $y = 0$, $z = 0$) and is the proxy for the maximum expansion. 

Compared to $\lambda_{2/1}$ that depends only on $D$ and $r_0$, $a_{y/x}$ also provides information on the different modes within the test sphere or for test spheres with different radial extents.

The y-component of the tidal field at the position $(D,r)$ in the reference frame of the host galaxy (see $a_y$ in Fig.~\ref{fig:tides_general}) is given by\footnotemark[2]:

\begin{equation} \label{Eq:ay}
|a_y| = \frac{1}{\sqrt{1+\left ( \frac{D}{r} \right )^2}} \cdot \frac{\partial \Phi(R)}{\partial R}
\end{equation}

\noindent
where $R = \sqrt{D^2+r^2}$ is the distance of $(D,r)$ to the host galaxy. For an NFW potential, $a_{y/x}$ is therefore: 

\begin{equation}\label{Eq:ayax}
\begin{split}
&	a_{y/x} (r, D, r_0) =\\
&\frac{\left(\sqrt{r^2+D^2}+r_0\right)  \log \left(\frac{\sqrt{r^2+D^2}+r_0}{r_0}\right)-\sqrt{r^2+D^2}}{\left(r^2+D^2\right) \sqrt{\frac{D^2}{r^2}+1} \left(\sqrt{r^2+D^2}+r_0\right) (A+B)}
\end{split}
\end{equation}

\noindent
with:

\begin{eqnarray}
A &=&  \frac{(-r+D+r_0) \log \left(\frac{-r+D+r_0}{r_0}\right)+r-D}{(r-D)^2 (-r+D+r_0)} \\
B &=& \frac{D-(D+r_0) \log \left(\frac{D+r_0}{r_0}\right)}{D^2 (D+r_0)}
\end{eqnarray}

\noindent
For a scale radius $r_0$ of the external NFW potential, the regimes where the compression $a_y$ is larger than the expansion $a_x$ are defined by $a_{y/x} > 1$. Fig.~\ref{Fig:ayax} illustrates these regions for scale radii $r_0$ between 10 and 100 kpc. For larger values of $r_0$ and therefore more massive host galaxies, the regions where compression dominates get bigger, they extend out to both larger distances $D$ as well as to larger radii $r$. 

The virial mass $M_{\mathrm{vir}}$ for a DM halo is defined as

\begin{equation}\label{Eq:Mvir}
M_{\mathrm{vir}} = 4 \pi \rho_0 r_0^3 \left ( \log(1+c) - \frac{c}{1+c}\right )  
\end{equation}

\noindent
with 

\begin{equation}\label{Eq:rho0}
\rho_0 = \rho_{\mathrm{cr}} \delta_{c}
\end{equation}

\noindent
and

\begin{equation} \label{Eq:deltacr}
\delta_{c} =\frac{200}{3}\frac{c^3}{\log(1+c) - c/(1+c)}
\end{equation}

\noindent
where $\rho_{\mathrm{cr}}= 3 H_0/(8 \pi G)$ is the critical density of the Universe, while $\delta_c$ is the characteristic over-density of the individual DM halo and dependent on the concentration parameter $c$. 
\citet{Correa2015} derive the following fitting functions for the $c-M_{\mathrm{vir}}$ relation (see Sec.~\ref{Sec:discussion} for a discussion) in the Planck cosmology:

\begin{equation}\label{Eq:cMrelation}
\log_{10} c = \alpha + \beta \log_{10} (M/\Msun) \left [  1 + \gamma \log_{10}^2 (M/\Msun) \right ]
\end{equation}

\noindent
with $\alpha=1.49809$, $\beta= -0.02499$, and $\gamma =  0.0056$  for z~=~0.

As $\rho_c$ and $c$ depend on $M_{\mathrm{vir}}$ in a non-trivial way, we iteratively calculate $M_{\mathrm{vir}}$ for the $r_0$ values shown in Fig.~\ref{Fig:ayax}. We choose $H_0 = 67.8\,\mathrm{km\,s^{-1}\,Mpc^{-1}}$ \citep{PlanckCollaboration2015} and tabulate the virial masses in a table on the right hand side of Fig.~\ref{Fig:ayax} for reference.

\begin{figure*}
	\includegraphics[width = 0.8\linewidth, bb = 0 10 310 170, clip]{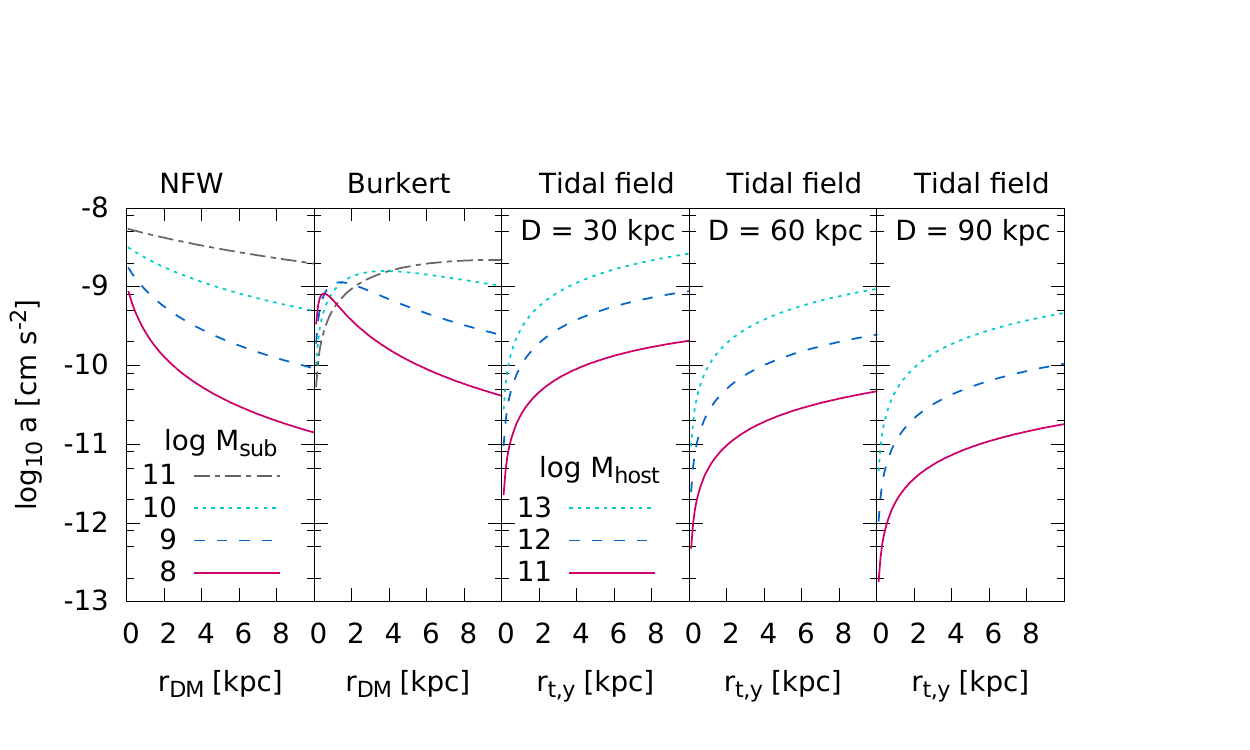}
	\caption{Comparison of the accelerations inside a DM sub-halo with an NFW (first panel) and a Burkert profile (second panel) as well as those in a tidal field at distances 30 (third panel), 60 (fourth panel), and 90 kpc (fifth panel) to the host galaxy. The accelerations are indicated for different sub-halo virial masses $M_{\mathrm{sub}}$ (first two panels) and host masses $M_{\mathrm{host}}$ (last three panels, see legend). We assume an NFW potential for the host galaxy in the case of the tidal field. Unlike the DM potential, the tidal field is not spherically symmetric around $r = 0$. We use here the compressing acceleration $a_y$ as indicated in Fig.~\ref{fig:tides_general} along the y-axis. In order to highlight the asymmetry the radius for the tidal field is labelled as $r_{t,y}$. }
	\label{fig:acc_comp}
\end{figure*}

\section{Comparison to dark matter sub-halos} \label{Sec:ayar}

In the previous section we have shown at which positions a tidal field can lead to a compression in y- and z-direction that is larger than the expansion it causes in x-direction. In its astrophysical application, this additional force can enhance star formation but could also stabilize the object against stellar feedback, similar to a DM sub-halo. 

Here we compare the accelerations inside a DM sub-halo $a_r = |\partial \Phi / \partial r| $ to the compressing acceleration caused by the tidal field $a_y$ (Eq.~\ref{Eq:ay}). Pure N-body simulations predict NFW profiles for DM halos on all scales, but observations show indications for centrally cored density profiles for DGs \citep[e.g.][]{DeNaray2008,Battaglia2008,DeBlok2010,Walker2011,Amorisco2012}.
Baryonic processes such as stellar feedback and a resulting wind are proposed as a mechanism to produce DM profiles with a central core \citep[see e.g.][]{Pontzen2012} but its realization is not yet fully understood \citep{Recchi2013}. In this section we compare the tidal field inside a massive NFW halo to DM sub-halos with both cuspy (NFW-NFW) and cored central regions (NFW-Burkert). As an example for a cored DM distribution, we use the Burkert halo, an empirical profile from \citet{Burkert1995}:

\begin{equation}\label{Eq:burkert}
\rho_{\mathrm{Burkert}}(R) = \frac{\rho_0 r_0^3}{(R + r_0) (R^2+r_0^2)} 
\end{equation}

\noindent
with the central density $\rho_0$ and a scale radius $r_0$.

The accelerations within each DM profile are shown in the first two panels of Fig.~\ref{fig:acc_comp} for sub-halo masses of $10^8$, $10^9$, $10^{10}$, and $10^{11}\,\Msun$. The third to fifth panel of Fig.~\ref{fig:acc_comp} demonstrate the accelerations along the y-axis (see Fig.~\ref{fig:tides_general}) for host masses of $10^{11}$, $10^{12}$, and $10^{13},\Msun$ for distances of 30, 60, and 90 kpc. 

Similar to $a_{y/x}$, we can compare the accelerations $a_{\mathrm{t,y}}(y)$ to $a_{\mathrm{DM}}(r)$, and find regimes where the tidal acceleration dominates and therefore $a_{t/DM} = a_{\mathrm{t,y}}(y)/a_{\mathrm{DM}}(r) \ge 1$

For each distance $D$ to the host galaxy and for each radius $r$ of a test sphere in this tidal field, the compressing tidal force can be related to the acceleration within a DM sub-halo with either a central cups (NFW, Sec.~\ref{Sec:NFWNFW}) or core (Burkert, Sec.~\ref{Sec:NFWBurkert}).

\subsection{NFW tidal field compared to an NFW sub-halo} \label{Sec:NFWNFW}

We are comparing here the compressing tidal accelerations of an NFW host halo $a_y$ to the gravitational accelerations inside an NFW sub-halo $a_r$:

 \begin{equation}
 |a_y| =  \frac{4 \pi G \rho_{0,t} r_{0,t}^3}{\sqrt{1+\left ( \frac{D}{r} \right )^2}}  \frac{(r_{0,t}+R)\log\left( \frac{r_{0,t}+R}{r_{0,t}} \right ) - R}{R^2(r_{0,t}+R)}
  \end{equation} 

\noindent
where $R = \sqrt{D^2+r^2}$ as in the case of $a_{y/x}$ (Eq.~\ref{Eq:ayax}) and

\begin{equation}
|a_r| = 4 \pi G  \rho_{0,d} r_{0,d}^3 \frac{(r_{0,d}+r) \log \left( \frac{r_{0,d}+r}{r_{0,d}} \right ) - r}{r^2 (r_{0,d}+r)} \quad .
\end{equation}

\noindent
$\rho_{0,t}$ and $r_{0,t}$ describe the potential of the host galaxy for the tidal field while $\rho_{0,d}$ and $r_{0,d}$ are the characteristic density and the scale radius of the DM sub-halo. 
The ratio $a_{y/r} = |a_y / a_r|$ between the tidal compression and a DM sub-halo can be calculated with\footnotemark[2]:

\begin{equation} \label{Eq:ayar}
a_{y/r} = \frac{\rho_{0,t} \; r_{0,t}^3 }{\rho_{0,d} \; r_{0,d}^3 } \; \frac{1}{ \sqrt{\left( \frac{D}{r}\right )^2 +1} } \; \frac{A}{B}   
\end{equation}

\noindent
where
\begin{eqnarray}
A &=& r^2(r+r_{0,d}) \left((R+r_{0,t}) \log \left(\frac{R+r_{0,t}}{r_{0,t}}\right)-R\right) \\
B &=& R^2 (R+r_{0,t}) \left((r+r_{0,d}) \log \left(\frac{r+r_{0,d}}{r_{0,d}}\right)-r\right)
\end{eqnarray}

 \begin{figure*}
 	\begin{center}
 		\includegraphics[width = 0.8\linewidth, bb = 50 12 275  205, clip] {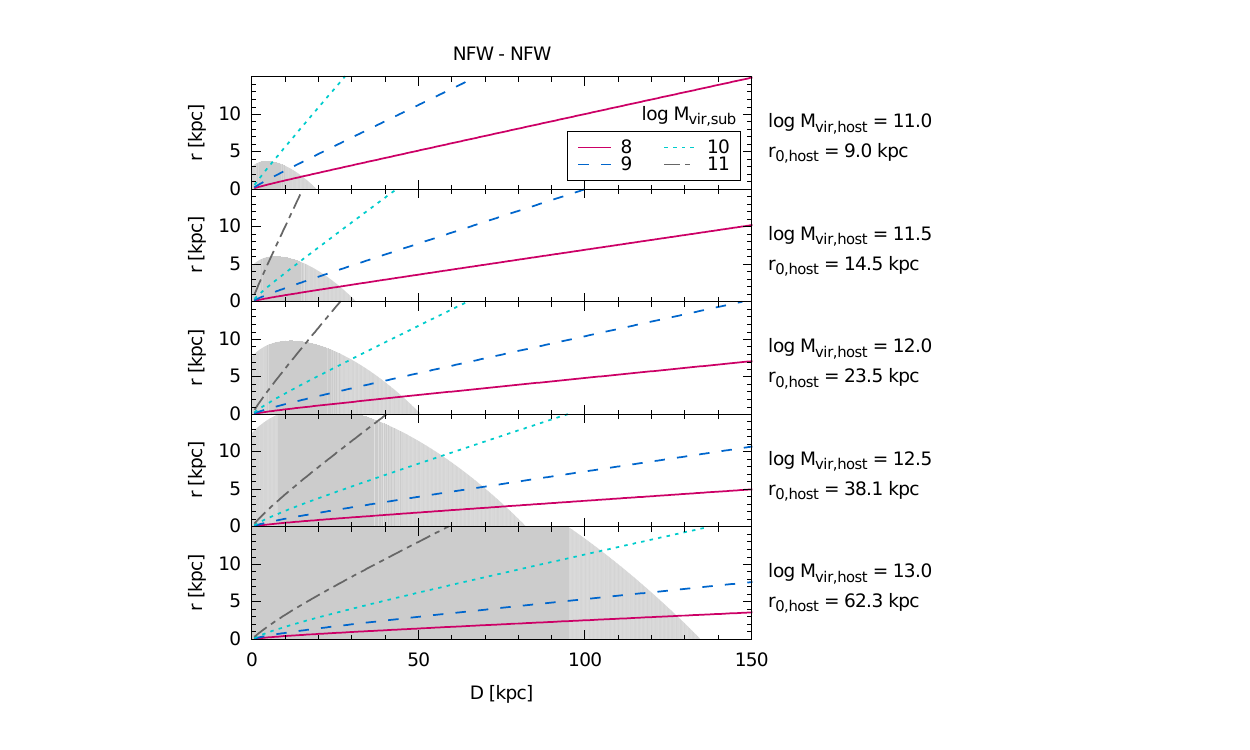}		
 	\end{center}
 	\caption{Comparison between the accelerations inside NFW DM sub-halos with virial masses $M_{\mathrm{vir,sub}}$ of $10^{8}$ (red solid lines), $10^{9}$ (blue dashed lines), $10^{10}$ (green dotted lines), and $10^{11}\,\Msun$ (grey dash-dotted lines) and in a tidal field of a host galaxy with virial masses $M_{\mathrm{vir,host}}$ from $10^{11}$ to $10^{13}\,\Msun$ (top to bottom panel). All lines represent the border of the region above which the tidal field is stronger than the representative DM sub-halo, therefore where $a_{\mathrm{t/DM}} = 1$ (Eq.~\ref{Eq:ayar}). For larger radii $r$ or smaller distances to the host $D$, the tidal field dominates.  The grey shaded area represents the region where the compressing acceleration in y direction is larger than the expansion in x-direction ($a_{y/x} > 1$, compare Fig.~\ref{Fig:ayax}).}
 	\label{fig:ayar}
 \end{figure*}

Assuming the relation between the concentration parameter $c$ and the virial mass (Eq.~\ref{Eq:cMrelation}) from \citet{Correa2015} allows to fully describe each halo profile with only one parameter, here the virial mass $M_{\mathrm{vir}}$. The characteristic density $\rho_0$ follows then from Eq.~\ref{Eq:rho0} and \ref{Eq:deltacr}, while the scale radius can be calculated by solving Eq.~\ref{Eq:Mvir} for $r_0$.
 
The relation $a_{y/r}$ therefore depends on $r$, $D$, and the virial masses of both the host galaxy and the comparison DM sub-halo. Fig.~\ref{fig:ayar} shows the lines for $a_{y/r} = 1$ for various host and sub-halo masses. The lines in each panel represent the edge of the regime where the compression of the tidal field is stronger than the DM sub-halo it is compared to. As the tidal field increases with larger $r$ while the acceleration inside an NFW sub-halo decreases at larger radii (see Fig.~\ref{fig:acc_comp}), the area above each line (for $r > r(a_{y/r} = 1)$) represents the regime where the tidal field dominates.  

An example: the compressing tidal field at a distance of $D = 100$ kpc to the center of a host halo of $M_{\mathrm{vir,host}} = 10^{13}\,\Msun$ (Fig.~\ref{fig:ayar}, bottom panel), is equivalent to the acceleration of a DM sub-halo of $M_{\mathrm{vir,sub}} = 10^8\,\Msun$ at a radius of $r = 2.5$ kpc. At larger radii, the tidal field gets stronger and it therefore compares to a DM sub-halo of $M_{\mathrm{vir,sub}} = 10^9\,\Msun$ at a radius of 5.5 kpc and $M_{\mathrm{vir,sub}} = 10^{10}\,\Msun$ at $r$ = 11 kpc. For applications and interpretations see Sec.~\ref{Sec:application}.

\begin{figure*}
	\begin{center}
		\includegraphics[width = 0.8\linewidth, bb = 50 12 275  205, clip] {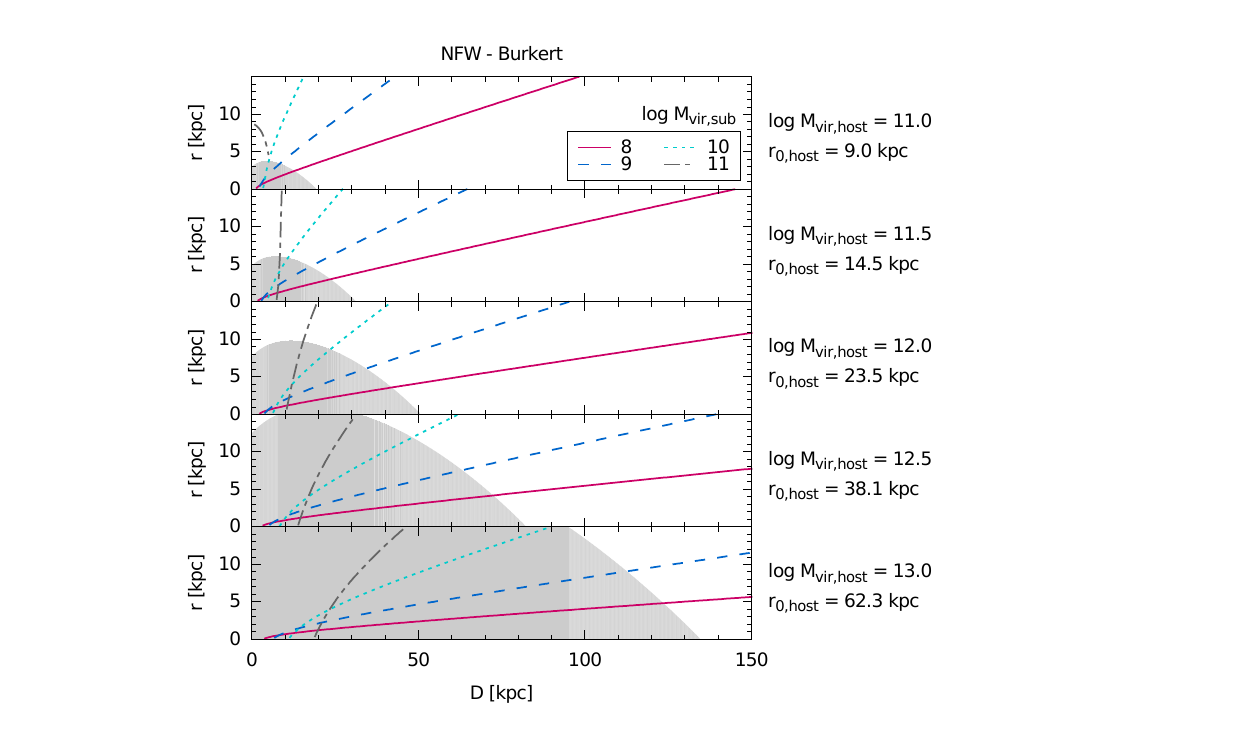}		
	\end{center}
	\caption{As Fig.~\ref{fig:ayar}, but here the DM sub-halos are represented with a cored potential from \citet{Burkert1995}.}
	\label{fig:ayar_burkert}
\end{figure*}

\subsection{NFW tidal field compared to a Burkert sub-halo} \label{Sec:NFWBurkert}

Analogously to Sec.~\ref{Sec:NFWNFW}, we repeat the same analysis for DM sub-halos with a Burkert profile. As we assume an NFW profile for the massive host galaxy, $a_y$ is still represented by Eq.~\ref{Eq:ay}. The radial acceleration inside a Burkert halo is given by:

\begin{equation}
\begin{split}
a_r &= \frac{\pi G \rho_{\mathrm{const}}  r_0^{7/3}}{r^2}  \times \\
& \left ( 2 \arctan \left ( \frac{r}{r_0} \right ) + 4 \log (r_0) - 2\log (r+r_0) - \log (r^2 + r_0^2)   \right )
\end{split}
\end{equation}

\noindent
where $\rho_0$ is the central DM density, $\rho_{\mathrm{const}}$ a constant that includes all constant factors for $\rho_0$ in cgs units (see Eq.~\ref{Eq:rhoconst}), and $r_0$ is the scale radius of the Burkert halo (Eq.~\ref{Eq:burkert}). According to \citet{Burkert1995}, they correlate with each other and the virial mass $M_{\mathrm{vir}}$ via:

\begin{eqnarray}
\rho_0 &=& 4.5 \times 10^{-2} \left [ \frac{r_0}{\mathrm{kpc}}  \right ]^{-2/3} \,\Msun\,\mathrm{pc}^{-3} \\ \label{Eq:rhoconst}
&=& \rho_{\mathrm{const}} r_0^{-2/3} \,\mathrm{g\,cm^{-3}} \\ 
M_{\mathrm{vir}} &=& 5.8 \times 7.2 \times 10^7 \times \left [ \frac{r_0}{\mathrm{kpc}}  \right ]^{7/3} \,\Msun
\end{eqnarray}

\noindent
The results for $a_{y/r} = 1$ are shown in Fig.~\ref{fig:ayar_burkert}.

\section{Applications} \label{Sec:application}

Cosmological simulations are now able to resolve dwarf galaxies with unprecedented details where effects of the variations of stellar feedback processes are studied carefully by different groups, as e.g. the EAGLE \citep{Schaye2014c}, the FIRE \citep{Hopkins2014b}, or the Illustris project \citep{Vogelsberger2014}. All of these objects are DM-dominated within $\Lambda$-CDM.

The data in both simulations and observations on young TDGs is very limited compared to the DM dwarf galaxies. Here, we relate the properties of the tidal field around TDGs to those of the DM halo of isolated dwarfs. This way it is possible to estimate the fate of TDGs as well as explain the peculiarity of the apparent DM content in some of the TDGs. 

This exercise assumes perfect NFW potentials for both the host galaxy as well as the DM sub-halo. This has to be taken with caution, especially as the potential of the host is time-variant during the galaxy interaction. Nevertheless it provides a quick and easy check on the acceleration regime in which the TDG is located, and allows first estimates on the potential survivability of TDGs.

\subsection{NGC 5557}

\citet{Duc2014} found three TDGs around the early-type galaxy NGC5557. SED fitting of the TDG NGC5557-E1 reveals an age of 4 Gyr, which makes it the oldest TDG uniquely identified so far. The work of this paper provides a possible explanation for the long-term survival of this object.

\citet{Cappellari2013a} derived a maximum value of the circular velocity for NGC5557 of $v_{\mathrm{max}} = 340\,\mathrm{km\,s}^{-1}$ from their best-fitting mass-follow-light JAM (Jeans Anisotropic Multi-gaussian expansion) model. The virial mass of NGC 5557 can therefore be estimated with

 \begin{equation} \label{Eq:vir}
 M_{\mathrm{vir} } = 2.33 \times 10^5 
 \left(\frac{v_{\mathrm{max}}}{\mathrm{km\,s}^{-1}} \right)^3 \,\Msun
 \end{equation}

\noindent 
for an NFW potential \citep{Navarro1998}. For the TDG NGC5557-E1 the host mass is therefore in the order of $\log_{10} M_{\mathrm{v,NGC5557}} = 13$. This is a very crude mass estimate, but to illustrate the application of this method, an order of magnitude approximation is enough.

From the images in \citet{Duc2014}, the projected distance of the TDG to its host is about $D \approx 70\,\mathrm{kpc}$ in projection and the HI contours show a spatial extent between $r \approx 5\,\mathrm{kpc}$ (minor axis) and $r \approx 6 \,\mathrm{kpc}$ (major axis). The combination of $\log M_{\mathrm{vir,host}} [\Msun] \approx 13$, $D \ge 70\,\mathrm{kpc}$, and $r = 0 - 6\,\mathrm{kpc}$ can be overlaid to Fig.~\ref{fig:ayar} which is shown as a short black vertical line in Fig.~\ref{fig:app}.

At the edge of TDG NGC 5557-E1 \citep[r $\approx$ 6 kpc][]{Duc2014} the accelerations are comparable to those of an NFW DM sub-halo with a few times $10^9\,\Msun$ or to a Burkert sub-halo with $10^9\,\Msun$. 
 Around the effective radius of NGC 5557-E1 ($r_e = 2.3\,\mathrm{kpc}$), where the baryonic matter is expected to dominate in all cases \citep{Toloba2011}, the additional acceleration from the tidal field is  still the same as in DM sub-halo with $M_{\mathrm{vir,sub}} \approx 10^{8}\,\Msun$. In addition, TDG NGC 5557-E1 is also well in the regime where the compression in y and z direction dominates the tidal stretching (grey shaded area in Fig.~\ref{fig:app}). 

This might be an explanation for the extended lifetime of this particular TDG. Detailed kinematical studies of NGC 5557 and its surroundings, used for a DM halo profile fit, would allow for a much better estimate. In principle, once the gravitational potential of the host galaxy is known, these regions can be constructed easily. 

As an estimate for the further evolution of this object, the tidal radius can be estimated with 

\begin{equation}\label{Eq:rtidal}
	r_j = \left( \frac{M_{\mathrm{TDG}}}{3M_{\mathrm{host}}(D)}\right)^{1/3} D
\end{equation}

\noindent
\citep{Binney2008}, where $M_{\mathrm{host}}(D)$ is the enclosed mass of the host halo up to the distance $D$. For NGC 5557-E1 with a mass of $M_{\mathrm{TDG}} = 1.2 \times 10^8\,\Msun$ and at a distance of 70 kpc, the $r_j$ is at 2 kpc, close to the current $r_e$.

As the estimates on the mass and DM profile of NGC 5557 are only based on one measurement of the maximum rotation velocity, no precise statements can be made yet about the individual objects. But some general conclusions for a TDG at a distance of 70 kpc to a host galaxy with a virial mass of $10^{13}\,\Msun$ and an NFW profile can be drawn nevertheless:

\begin{figure*}
	\begin{center}
		\includegraphics[width = 0.9\linewidth, bb = 50 12 275 100, clip]{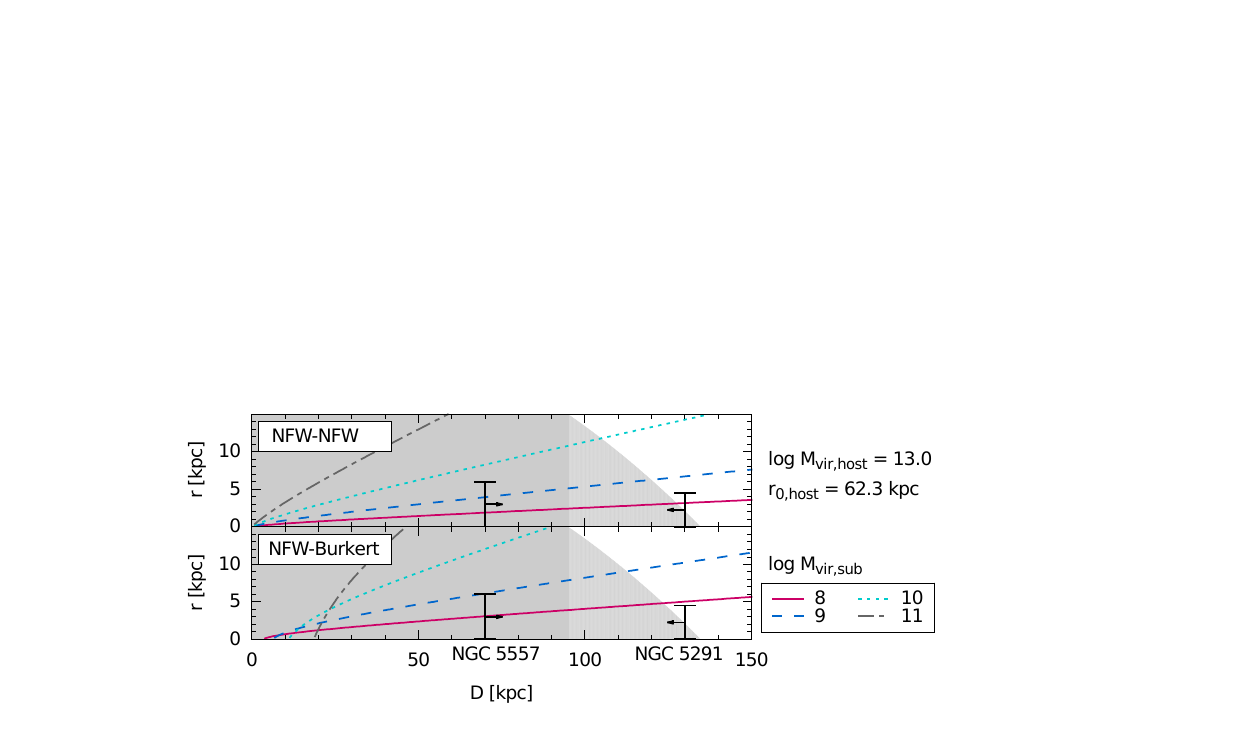} 				
	\end{center}
	\caption{Labels as in Fig.~\ref{fig:ayar}. In addition the estimated distances of the TDGs around NGC 5557 and NGC 5291 are indicated as black vertical lines at D = 70 kpc (NGC 5557-E1) and D = 130 kpc. For the NGC 5557 system, D is the projected distance and therefore a lower limit, as indicated by an arrow. \citet{Bournaud2007} performed simulations of the galaxy interaction that caused the tidal features around NGC 5291 and their best model places the TDGs at a distance of around 130 kpc today. Because of the special geometry of the interaction, they spent most of their evolution closer to the host galaxy than it is observed today. This is indicated by an arrow towards smaller D. The top panel shows the case NFW-NFW (Sec.~\ref{Sec:NFWNFW}) and the bottom panel the case NFW-Burkert (Sec.~\ref{Sec:NFWBurkert}).}
	\label{fig:app}
\end{figure*} 

\begin{itemize}
	\item[--] Around 5 kpc away from the center of the TDG, gas from the surrounding tidal arm is compressed by the tidal field onto the TDG, as if the TDG was embedded in a DM halo of a few times $10^{9}\,\Msun$.
	\item[--] This effect increases with increasing radii ($10^{10}\,\Msun$ at 11 kpc for the NFW-NFW case), which is an efficient way to bring gaseous material close to the centre of the TDG. 
	\item[--] The tidal field can fuel SF and stabilize the outer parts of the TDG against stellar feedback, just as a DM halo in this mass range could. 
	\item[--] In the very central region of the TDG, the contribution of the tidal field vanishes. But also DM-dominated DGs are dominated by the self-gravity of the baryonic matter close to their centre. 
	\item[--] At this distance (70 kpc), the TDGs are clearly in the regime where the compression of the tidal field dominates the tidal stretching ($a_{y/x} > 1$).  
	\item[--] For a TDG with $10^8\,\Msun$ the dynamics at the outer parts could mimic a DM content in the order of a few times the baryonic mass, caused by the additional acceleration of the tidal field.  
	\item[--] An estimate of the tidal radius gives $r_j = 2\,\mathrm{kpc}$. Even if all the gas is blown out eventually, a bound stellar object with a radius of around 2 kpc could survive, if it's on a circular orbit.  
\end{itemize}

\subsection{NGC 5291}

\citet{Bournaud2007} measured the rotation curves of three TDGs in the gaseous ring of NGC 5291. They found dynamical masses that are between two and three times larger than their visible masses. TDGs cannot contain bound DM particles as they originate from disk material and they are not massive enough to capture the high velocity dispersion particles from their host galaxies. If the discrepancy between the visible and dynamical mass is real and not caused by e.g. projection effects, it has to be explained otherwise. Among the proposed solutions are a modified Newtonian gravity \citep{Gentile2007}), a molecular gas content that is 10 times larger than expected with the traditional CO to molecular gas conversion factor \citep{Bournaud2007}, or a velocity anisotropy caused by the tidal field \citep{Kroupa1997}.

NGC 5291 is included in the HIPASS (HI Parkes All Sky Survey) database and \citet{Koribalski2004} measured a 50\% velocity line width of $w_{50} = 637\,\mathrm{km\,s}^{-1}$. Assuming that the maximum rotation speed $v_{\mathrm{max}}$ is close to $0.5 \times w_{50}$, the virial mass of NGC 5291 is estimated to be $M_{\mathrm{v,NGC5291}} = 7.5 \times 10^{12}\,\Msun$ (Eq.~\ref{Eq:vir}) or $\log_{10} M_{\mathrm{v,NGC5291}} = 12.9$.

Simulations of galaxy interactions to create the morphology of the gaseous ring around NGC 5291 favour a scenario of a high speed collision between NGC 5291 and the massive elliptical galaxy IC 4329 around 360 Myr ago \citep{Bournaud2007}. Their best model has the ring located in a plane, 130 kpc away from the host galaxy today. 

For an estimate on the future of the TDGs around NGC 5291, they are indicated in Fig.~\ref{fig:app} at $D = 130\,\mathrm{kpc}$. Their rotation curves were measured out to 5 kpc \citep{Bournaud2007}, where their acceleration by the tidal field is comparable to that of a DM sub-halo with around $10^8\,\Msun$. This could account for some of the missing mass for the TDGs in NGC 5291. 

As the ring was produced by an high speed encounter, a significant part of the evolution of the TDG was closer to the host galaxy (indicated with an arrow in Fig.~\ref{fig:app}). Nevertheless, the TDGs today are on the edge of the regime where $a_{y/x}<1$ and therefore where the expansion dominates the contraction. If they are not on eccentric orbits that will bring them closer to the host galaxy, they are more likely to get disrupted. 

As in the case of NGC 5557, only little is known about the detailed DM profile of NGC 5291. We assume again an NFW potential, but if the disturbance that caused the tidal ring has happened only a few hundred Myr ago, the DM halo of NGC 5291 is very likely not re-virialised. In addition, the estimates of $M_{\mathrm{v,NGC5291}}$ and the distances are first order approximations and should be used with great care. More detailed data on the DM distribution from simulations or observations would allow stronger statements on their future survivability. 

In general, the following information can be extracted for a TDG at a distance of 130 kpc to a $10^{13}\,\Msun$ host galaxy with an NFW DM halo profile:

\begin{itemize}
	\item[--] At $D$ = 130 kpc, the tidal field is much weaker than in the previous example with $D$ = 70 kpc. 
	\item[--] The tidal radius (Eq.~\ref{Eq:rtidal}) for a an object with $5\times10^8\,\Msun$ is $r_j = 5\,\mathrm{kpc}$. 
	\item[--] The TDG is located at the edge of the $a_{y/x} > 1$ region. At a radius of 5 kpc, the tidal stretching in x-direction is even stronger than the compression in y- and z- direction. In this regime, the formation of TDGs is not enhanced by the tidal field, as less mass can be channelled from the tidal arm to the close vicinity of the TDG.  
	\item[--] The compressing accelerations from the tidal field are comparable to those of a $10^8\,\Msun$ DM halo, therfore the dynamical stage of the TDG could mimic a low-mass $\approx 10^8\,\Msun$ DM halo.
	\item[--] If the TDG moves to larger distances, the tidal radius would increase while the compressing tidal field decreases. An object like that would be unsupported by any stabilizing extra acceleration field and if it is still gas-rich at this stage, stellar feedback may be efficient in expelling the gas.  
\end{itemize}

If the gravitational potential of NGC 5291 has a higher concentration parameter $c$
than the average value of \citet{Correa2015} the tidal field is stronger, resulting in
a higher dynamical mass in its TDGs. In order to explain an additional 
dynamical mass of $\approx 2 \times 10^9\,\Msun$, as measured in  \citet{Bournaud2007} 
for the TDGs around NGC 5291, the concentration parameter of the DM halo of NGC 5291 would have to be unrealistically large ($\gg$ 1000). 

Note, that rather than making clear predictions on individual TDGs, those objects were chosen to illustrate which information can be extracted with this simple approach.

\section{Discussion} \label{Sec:discussion}

We have presented a simple analytic method to investigate the survivability of TDGs by comparing the accelerations caused by the tidal field to those inside a DM sub-halo. In this section we discuss the limitations of the presented approach.

In general, the compression due to the tidal field increases with decreasing distance to the host galaxy (Eq.~\ref{Eq:gauss}). Using our terminology, this translates into higher equivalent DM sub-halo masses and therefore an increased stability of the TDG against stellar feedback processes. The lower limit on the distance between TDG and host galaxy is considered by a calculation of the tidal radius. While the compression increases with decreasing distance, the volume within which test particles are gravitationally bound to the satellite object, rather than to the host galaxy, decreases. This volume is characterised by the tidal (or Jacobi) radius (Eq.~\ref{Eq:rtidal}). We do not directly include a lower distance limit in the method here, but we compare the tidal radius to the physical extent of the object for the presented applications as an independent test. 

Another process that limits the survivability of TDGs with smaller distances $D$ to the host galaxy is the orbital decay caused by dynamical friction \citep{Chandrasekhar1943}. Simulations of interacting galaxies have shown that TDGs that form close to the host galaxy ($<50\,\mathrm{kpc}$) can be re-accreted on a time-scale of a few hundred Myr \citep[e.g.][]{Hibbard1995, Bournaud2006}.  Note, however, that the dynamical friction time-scale for a TDG is much longer than for a DM-dominated satellite galaxy with the same baryonic mass, as $\mathrm{d}\vec{v}_{\mathrm{orb}} /\mathrm{d}t $ is proportional to the total mass of the satellite \citep{Chandrasekhar1943}.

Summarizing, we expect a ``sweet spot" at particular distances to the host galaxy, where the survivability of TDGs is enhanced by the tidal field. The distance range depends on the mass and shape of the DM halo of the host and the analytic method presented here can help to determine this range. It also allows one to predict the survivability of individual TDGs given their distance to their host galaxies and the masses of their host halos, if the survivability of DGs in DM halos with equal accelerations is known. The long-term stability of a TDG is favoured if the distance to the host galaxy is sufficiently small for the tidal field to stabilize it against stellar feedback, but also large enough that its tidal radius does not get too small and dynamical friction will not lead to orbital decay on a short time-scale.

We made several assumptions and simplifications:

{\it Eccentric orbits:} The method presented here only depends on the current distance. To predict the further evolution or to reconstruct the past, the orbit and therefore the time-dependence of $D$ has to be taken into account. A highly eccentric trajectory will bring the TDG through very different regimes. Note however that long-living TDGs are expected to move preferentially on orbits with mild eccentricity as otherwise they would get too close to the host galaxy eventually.

{\it Host halo potential profile:} We assume spherically symmetric NFW halos for all host galaxies. At the late stages of a galaxy interaction, especially after the galaxies have merged, this assumption is expected to be valid. During the galaxy interaction event, the DM distribution is not virialised and might therefore deviate significantly from the NFW profile. Simulations can provide time-dependent gravitational potentials and the tidal compression can then again be compared to accelerations in DM sub-halos. 

{\it c-M relation:} For all NFW halos, including both the host galaxy and the DG, we assume the relation between the concentration parameter and the virial mass from \citet{Correa2015}. Their relation combines an analytic model for the halo mass accretion history based on the extended Press Schechter theory with a fit between the concentration parameter and the formation time from DM-only simulations. We consider the DM profiles of average host galaxies but real galaxies are expected to scatter around the c-M relation.

{\it Asymmetry of the tidal field:} In contrast to the gravitational field inside an isolated DM sub-halo, the tidal field is not spherically symmetric. When we compare the accelerations, we take the maximum compression of the tidal field $a_y$ (see Fig.~\ref{fig:tides_general}). The extension in the direction towards the centre of the host halo is accounted for when the ratio between compression and expansion is calculated. In the chosen geometry of this toy model, the tidal accelerations and eigenvalues of the tidal tensor in y-direction are equal to those in z-direction. In order to formulate a conservative criterion, we define the regions where the compression dominates the expansion with $|a_y/a_x|>1$ and $\lambda_2/\lambda_1 < -1$. If the z-direction, where the tidal field is as compressive as in the y-direction, is taken into account as well, the above conditions would be $|a_y/a_x|>0.5$ and $\lambda_2/\lambda_1 < -0.5$, which increases the distance range where the survival of TDGs is boosted.  

{\it Internal kinematics:} We neglect all internal kinematics, such as accelerations caused by self-gravity of the stellar and gaseous components, kinetic stellar feedback or rotation. All these processes exist and their significance may vary greatly. Therefore, we do not address the very inner part of both TDGs and isolated DGs and focus our analysis on the outer regimes where the DM or the tidal field dominates over the self-gravity of the baryons. Note that tidal compression is more important than tidal expansion, as it brings material close to the centre, where it can become trapped by the self-gravity of the baryonic matter. For the detailed study of the evolution of individual TDGs or DGs, computationally expensive, high-resolution simulations are necessary. The method proposed in this work allows a quick way to gauge the importance of a tidal field.  

{\it General remark:}
During the very early stages of the galaxy interaction, the rapidly changing total gravitational potential leads to regions of fully compressive tidal modes that are only short lived \citep{Renaud2009}. These regions can enhance the formation of bound structures. For later stages of the interaction, when the TDG is orbiting around the host galaxy, the method presented here can help to understand the role of the tidal field in the long-term survival. The two approaches are therefore complementary.

\section*{Acknowledgements}

I am grateful for helpful discussions and very appreciated comments from Joop Schaye, Gerhard Hensler, Bart Clauwens, Simone Recchi, Pavel Kroupa, Pierre-Alain Duc, and Florent Renaud. I acknowledge support from the European Research Council under the European Union's Seventh Framework Programme (FP7/2007-2013)/ERC Grant agreement 278594-GasAroundGalaxies.

\end{document}